\documentclass[12pt,preprint]{aastex}

\def\ltsima{$\; \buildrel < \over \sim \;$}
\def\gtsima{$\; \buildrel > \over \sim \;$}
\def\lsim{\lower.5ex\hbox{\ltsima}}
\def\gsim{\lower.5ex\hbox{\gtsima}}
\def\lapp{\ifmmode\stackrel{<}{_{\sim}}\else$\stackrel{<}{_{\sim}}$\fi}
\def\gapp{\ifmmode\stackrel{>}{_{\sim}}\else$\stackrel{<}{_{\sim}}$\fi}

\setbox0=\hbox{\rm0}
 
\shorttitle{Spectroscopic Screening of Ter5 giants}
\shortauthors{Origlia et al.}
 
\begin{document} 

\title{Spectroscopy Unveils the Complex Nature of Terzan~5
\footnote{Based on observations collected at the W.M.Keck
    Observatory, which is operated as a scientific partnership among
    the California Institute of Technology, the University of
    California, and the National Aeronautics and Space Administration.
    The Observatory was made possible by the generous financial
    support of the W. M. Keck Foundation.}}

\author{
L. Origlia\altaffilmark{2},
R. M. Rich\altaffilmark{3},
F. R. Ferraro\altaffilmark{4},
B. Lanzoni\altaffilmark{4},
M. Bellazzini\altaffilmark{2},
E. Dalessandro\altaffilmark{4},
A. Mucciarelli\altaffilmark{4},
E. Valenti\altaffilmark{5},
G. Beccari\altaffilmark{5}
}
\affil{\altaffilmark{2} INAF-Osservatorio Astronomico di Bologna,
Via Ranzani 1, I-40127 Bologna, Italy, livia.origlia@oabo.inaf.it}

\affil{\altaffilmark{3} Physics and Astronomy Bldg, 430 Portola Plaza
  Box 951547 Department of Physics and Astronomy, University of
  California at Los Angeles, Los Angeles, CA 90095-1547}

\affil{\altaffilmark{4} Dipartimento di Astronomia, Universit\`a degli Studi
di Bologna, via Ranzani 1, I--40127 Bologna, Italy}

\affil{\altaffilmark{5} ESO - European Southern Observatory,
  Karl-Schwarzschild Str. 2, D-85748 Garching bei M\"unchen,  Germany} 


\begin{abstract}
We present the chemical abundance analysis of 33 red giant stars
belonging to the complex stellar system Terzan~5.  We confirm the
discovery of two stellar populations (Ferraro et al. 2009, Nature,
462,483) with distinct iron abundances: a relatively metal-poor
component with [Fe/H]$= -0.25\pm0.07$ r.m.s., and another component with
[Fe/H]$=+0.27\pm0.04$ r.m.s., {\it exceeding in metallicity any known
 Galactic globular cluster}.  The two populations also show different
[$\alpha$/Fe] abundance ratios. The metal-poor component has an
average [$\alpha$/Fe]$= +0.34\pm 0.06$ r.m.s., consistent with the canonical
scenario for rapid enrichment by core collapse supernovae (SNe).  The
metal-rich component has [$\alpha$/Fe]$ = +0.03 \pm 0.04$ r.m.s., suggesting
that the gas from which it formed was polluted by both type II and
type Ia SNe on a longer timescale.  Neither of the two populations
shows evidence of the [Al/Fe] over [O/Fe] anti-correlation, that is
typically observed in Galactic globular clusters.  Because these
chemical abundance patterns are unique, we propose that Terzan 5 is
not a true globular cluster, but a stellar system with a much more
complex history of star formation and chemical enrichment.
\end{abstract}


\section{Introduction}
\label{intro}

Terzan~5 is commonly catalogued as a globular cluster (GC) located in
the inner bulge of our Galaxy. It is heavily reddened, with an average
color excess $E(B-V)=2.38$ \citep{bar98,val07} and such a reddening
strongly depends on the line of sight \citep{ort96,val07}.  This
stellar system also harbors an exceptionally large population of
millisecond pulsars (MSPs): indeed, the 34 MSPs detected so-far in
Terzan~5 amount to $\sim 25\%$ of the entire sample of known MSPs in
Galactic GCs \citep[][ see the updated list at {\tt
    http://www.naic.edu/$\sim$pfreire/GCpsr.html}]{ransom05}.

Recently, a combined photometric and spectroscopic study of Terzan~5
has led to the discovery of two distinct populations, as traced by two
well separated ($\delta K\simeq 0.3$) Red Clumps in the $(K,J-K)$
Color Magnitude Diagram (CMD), with a $\approx$0.5 dex difference in
their iron content \citep[][ hereafter F09]{fer09}.  A conventional
isochrone fit is consistent with the two populations of Terzan 5 being
separated by a few Gyr (F09), although only a small age gap is needed
if the younger population is enhanced in helium \citep{danto10}.

The findings in F09 appear to be best understood if Terzan~5 was much
more massive in the past than today, in order to retain the SN ejecta
and igniting other star formation episodes.  A more massive
proto-Terzan~5 would also naturally explain its large population of
MSPs and the fact that the metal-rich component is more centrally
concentrated than the metal-poor one \citep[F09; see also][]{lan10}, a
typical feature of stellar systems which are self-enriched in iron,
as, e.g., the dwarf galaxies.

With the aim of accurately reconstructing the puzzle of the formation
and evolutionary history of Terzan~5, we are currently undertaking a
global study of the photometric, chemical and kinematic properties
of its stellar populations.  This Letter presents the results of the
spectroscopic screening of a suitable sample of giant stars, in order
to obtain chemical abundances and abundance patterns of key metals,
like iron, carbon, aluminum, oxygen and other $\alpha$-elements, and
constrain the complex chemical enrichment history of Terzan~5.
 
\section{Observations and abundance analysis}

In order to select suitable targets, we used the
differential-reddening corrected optical-IR CMD of Terzan~5 shown in
Figure 4 of F09. In this diagram it is possible to recognize not only
two Red Clumps, but also two main red giant branches (RGBs). We
therefore selected a sample of red giants mostly located along the two
RGBs and spanning the entire luminosity range above the horizontal branch (HB) level, with
the purpose of fully characterizing their chemical content.  High
resolution spectra have been acquired on July 1-2, 2010 by using
NIRSPEC \citep{ml98} at Keck II.  A slit width of $0.43\arcsec$,
giving an overall spectral resolution R=25,000, and the standard
NIRSPEC-5 setting, which covers a large fraction of the 1.5-1.8 $\mu$m
$H$-band, have been used.

About 40 stars have been observed during the run. Here we report and
discuss the results for 33 giants having radial velocities consistent
with the systemic velocity of Terzan~5 \citep[see e.g.][]{harris,ori04,fer09}, 
i.e. likely members of the system.

\begin{figure*}[!hp]
\begin{center}
\includegraphics[scale=0.7]{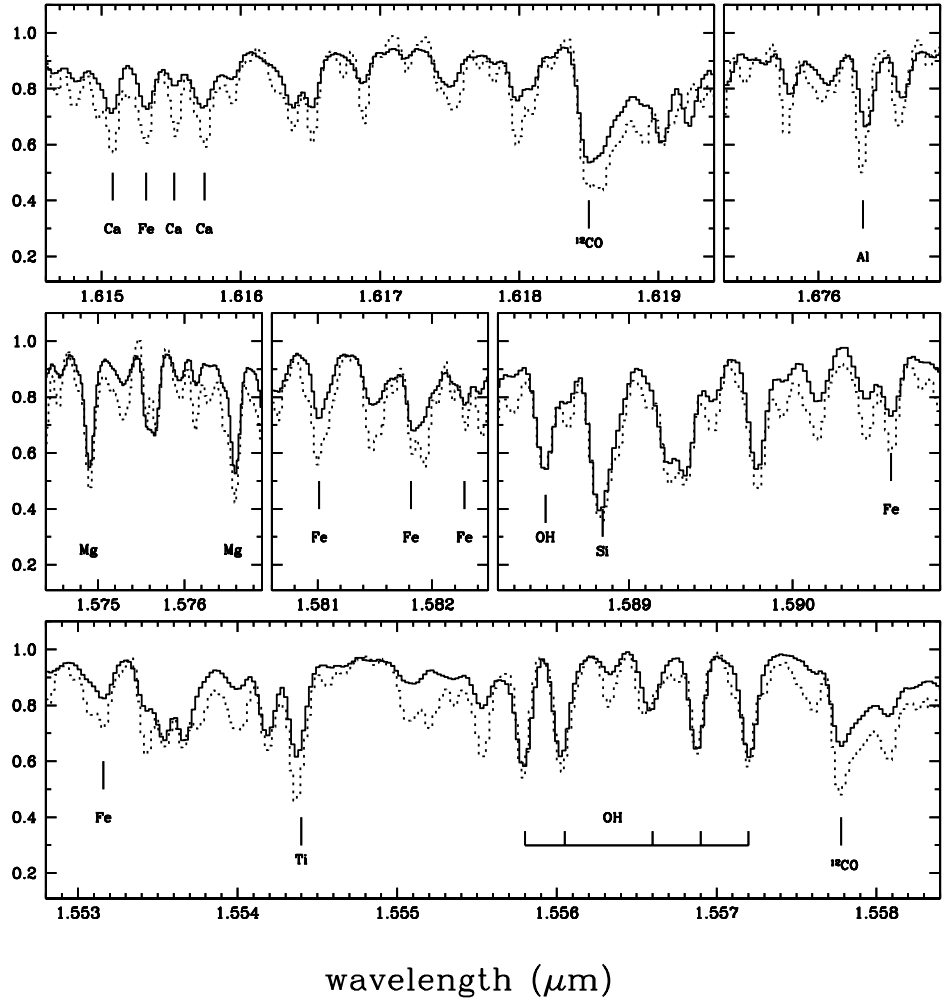}
\caption{Portion of the NIRSPEC $H$-band spectra of two red giants of
  Terzan~5 with similar $\rm T_{eff}\approx 3800$ K, but different
  chemical abundance patterns (solid line for the metal-poor star,
  dotted line for the metal-rich one).  A few atomic lines and
  molecular bands of interest are marked.}
\label{spec}
\end{center}
\end{figure*}

The raw spectra have been reduced using the REDSPEC IDL-based package
written at the UCLA IR Laboratory.  Each order has been sky subtracted
by using nodding pairs, and flat-field corrected.  Wavelength
calibration has been performed using arc lamps and a 2-nd order
polynomial solution, while telluric features have been removed by
using a O-star featureless spectrum observed during the same nights.
The signal to noise ratio of the final spectra is always larger than
30.  Figure~\ref{spec} shows an example of the observed spectra.

A grid of suitable synthetic spectra of giant stars has been computed
by varying the photospheric parameters and the element abundances, by
using an updated version of the code described in \citet{OMO93}.  By
combining full spectral synthesis analysis and equivalent width
measurements of selected lines, we have derived abundances for Fe, C,
O, Ca, Si, Mg, Ti and Al.  Our line list and details of the analysis
are given in \citet{orc02} and \citet{ori04}.  Reference Solar
abundances are from \citet{gv98}.

Stellar temperatures have been first estimated from the ($J-K$)$_0$
colors, by using the E($B-V$)=2.38 average reddening and the
color-temperature transformation by \citet{mon98}, specifically
calibrated on GC giants.  Gravity has been estimated from theoretical
evolutionary tracks, according to the location of the stars on the
RGB, while for the microturbulence velocity an average value of 2 km/s
has been adopted \citep[see][ and references therein for a detailed
  discussion]{ori97}.  More stringent constraints on the stellar
parameters have been obtained by the simultaneous spectral fitting of
several CO and OH molecular bands, which are very sensitive to
temperature, gravity and microturbulence variations \citep[see Figs. 6
  and 7 in][]{orc02}.  The final values of the adopted stellar
parameters, radial velocity and our best-fit chemical abundances with
$\rm 1\sigma$ random errors are listed in Table~\ref{tab1}.  We
conservatively estimate that the systematic errors in the derived
best-fit abundances, due to the residual uncertainty in the adopted
stellar parameters, are $\approx \pm 0.1$ dex.  However, it must be
noted that any variation in the stellar parameters makes all the
spectral features under consideration vary in a similar way (although
with different sensitivities). Hence, the derived {\em relative}
abundances are less dependent on the adopted stellar parameters,
i.e. they are affected by smaller systematic errors.

\section{Results and Discussion}

By the inspection of the iron abundance distribution (see
Table~\ref{tab1}), the existence of two populations is clearly
evident: a relatively metal-poor component (as traced by 20 giants in
our sample) with average [Fe/H]$= -0.25\pm 0.07$ r.m.s.,
and a metal-rich component (as traced by 13 giants in our sample) with average
[Fe/H]$=+0.27\pm 0.04$ r.m.s.. The two populations therefore show a $\rm
\Delta$[Fe/H]$\approx 0.5$ dex iron abundance difference, fully
confirming first results by F09 based on the observations of a small
sample of Red Clump stars.

\begin{figure*}
\begin{center}
\includegraphics[scale=0.7]{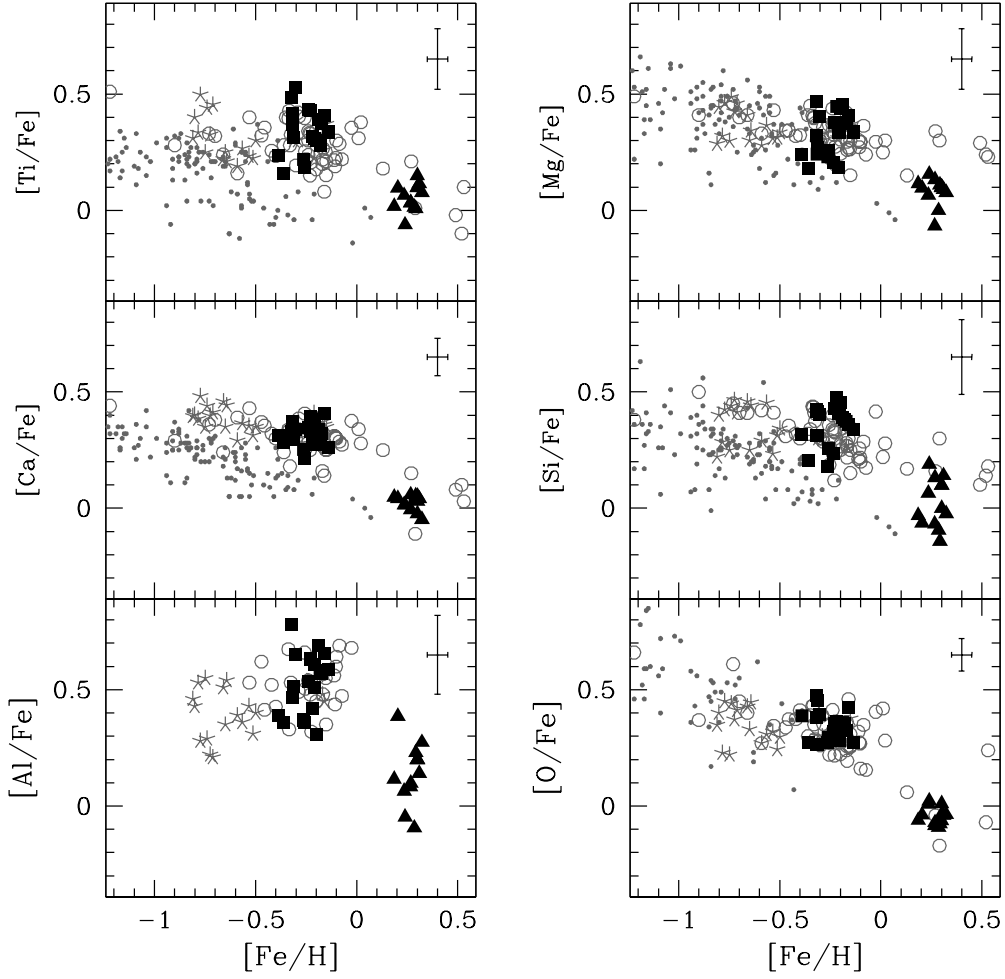}
\caption{Individual [$\alpha$/Fe] and [Al/Fe] abundance ratios as a function of
  [Fe/H] for the 20 metal-poor (solid squares) and the 13 metal-rich
  (solid triangles) giants in our observed sample. Included are also
  the four metal-poor giants observed by \citet{ori04}.  Typical
  errorbars are plotted in the top-right corner of each panel.  The
  abundance patterns of bulge field (open circles) and globular cluster (asterisks) giants
  \citep[see, e.g.,][and reference therein]{rich07,ful07,ori08} and
  those of disk and halo (gray dots) \citep{gra03} stars are also plotted for
  comparison.}
\label{alpha}
\end{center}
\end{figure*}

Figure~\ref{alpha} shows the various [$\alpha$/Fe] abundance ratios
for the giants observed in Terzan~5 and in other Galactic stellar
populations, for comparison.  The computed average abundance ratios
for the two Terzan~5 components are listed in Table~\ref{tab2}.  We
find an overall average [$\alpha$/H]$\approx +0.09$ and
[$\alpha$/Fe]$\approx +0.34$ for the metal-poor, and
[$\alpha$/H]$\approx +0.30$ and [$\alpha$/Fe]$\approx +0.03$ for the
metal-rich stars.  Hence the two populations show $\rm
\Delta[\alpha/H]\approx +0.2$ dex and a $\rm \Delta[\alpha/Fe]\approx
-0.31$ differences.

The first important result of this study is that  the two populations in Terzan~5
have clearly distinct abundance patterns, the most metal-rich being
significantly enriched in iron and moderately enriched in
[$\alpha$/H], with respect to the metal-poor one.  These properties
have never been observed before in any Galactic GC\footnote{Only
    the old, populous, disk open cluster NGC 6791 has iron and chemical
    abundance patterns \citep[e.g.][]{ori06} very similar to the
    Terzan~5 metal-rich population.}.  In fact, all GCs are
characterized by an extremely high homogeneity in the iron
abundance\footnote{Recent claims of the detection of much smaller
  ($\simeq$0.1 dex) iron dispersion have been reported in the
  literature for M22 \citep{marino09,dacosta09}, NGC 1851
  \citep{Car10a_1851}, and M54 \citep{Car10b_M54}.}.  The only notable
exception (but in a significantly lower metallicity regime) is the
halo globular-like system $\omega$ Centauri
\citep{norris95,soll05,JoPila}, which is, for this reason, currently
believed to be the remnant of an accreted and partially disrupted
dwarf galaxy \citep{bekki03}. Indeed, observational evidence of tidal debris 
from $\omega$ Centauri already exist \citep[see e.g.][and references therein]{wyl10}. 

The overall iron abundance and the [$\alpha$/Fe] enhancement of the
Terzan~5 metal-poor component is consistent with what is measured in
bulge stars (see Fig.~\ref{alpha}). This suggests that, as the bulk of
the bulge population, this component in Terzan~5 likely formed from a
gas mainly enriched by type II SNe on a short timescale.  When
compared to the metal-poor component, the larger [$\alpha$/H] overall
abundance of the metal-rich one can be explained with an additional
enrichment by type II SNe, whose ejecta must have been retained within
the potential well in spite of the violent explosions. Moreover, the
Solar [$\alpha$/Fe] abundance ratio indicates that its progenitor gas
was also polluted by type Ia SNe explosions on longer timescales.
Among globular cluster-like stellar systems, such a signature of 
type Ia SNe enrichment has been observed only in
the most metal rich population of $\omega$ Centauri \citep[see
e.g.][]{pan02,ori03,JoPila}.

Figure~\ref{alpha} also shows the [Al/Fe] {\it vs} [Fe/H] abundance distribution,
indicating that Aluminum behaves like $\alpha$-elements.  

\begin{figure}[h]
\begin{center}
\includegraphics[scale=0.7]{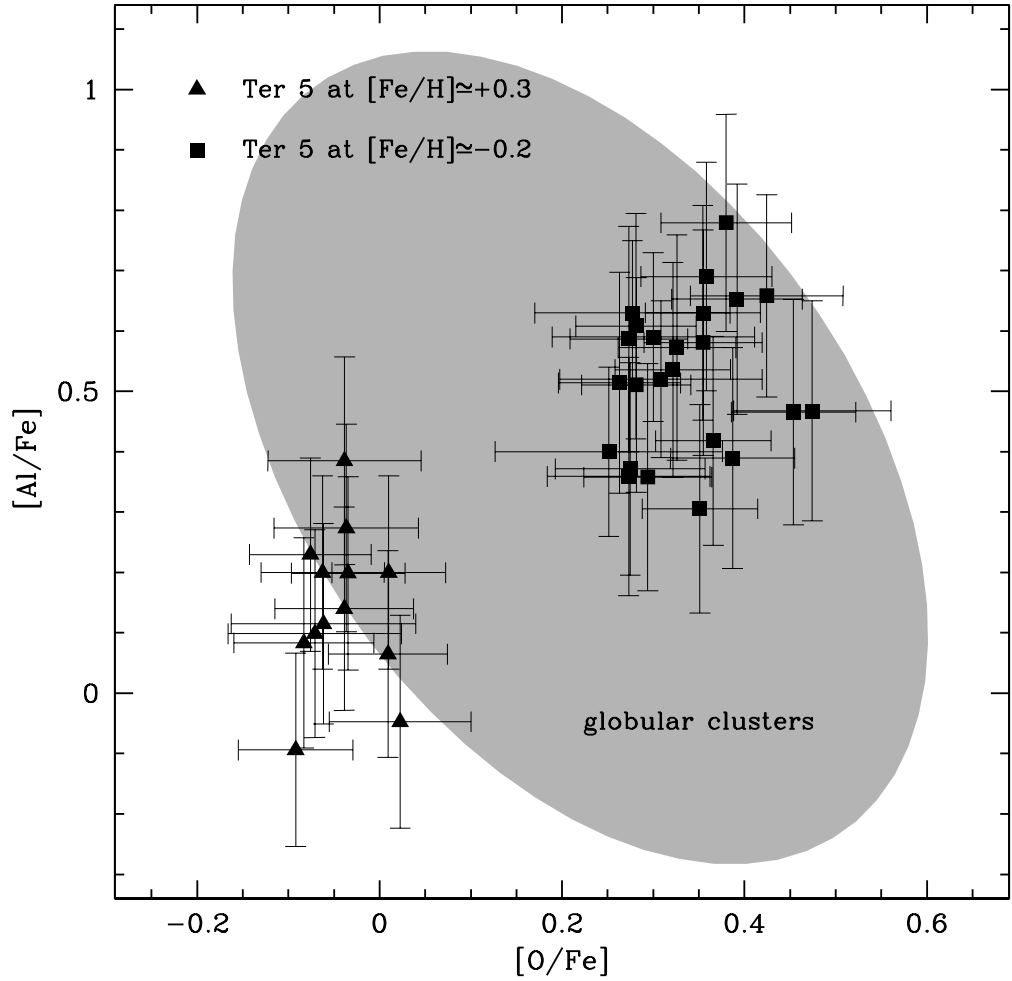}
\caption{[Al/Fe] {\it vs} [O/Fe] abundance ratios of the observed
  giants in Terzan~5.  Symbols are as in Fig.~\ref{alpha}.  The gray
  ellipse indicates the range of values measured in Galactic GCs
  \citep[see e.g.][]{Car10c}.}
\label{al}
\end{center}
\end{figure}

Figure~\ref{al} shows the [Al/Fe] {\it vs} [O/Fe] abundance ratios. 
As a whole, the stars of Terzan~5 display a clear
positive correlation between [Al/Fe] and [O/Fe], at variance with the
anti-correlation shown by ordinary GCs.  In fact, it is well known
that even genuine, single-metallicity GCs show large (up to $\sim 1$ dex)
star-to-star variations in the abundance of light elements (like Na,
O, Mg, and Al) that are not observed in the Galactic field stars, nor
in the field of nearby dwarf galaxies \citep[see,
  e.g.,][]{Car10b_M54}. In particular, [Na/Fe] and [Al/Fe] abundances
are seen to anti-correlate with [O/Fe] in all GCs that have been
surveyed till the present day, both in the Galaxy and beyond
\citep[see][and references therein]{leta,Car09,muccia}.  This chemical
fingerprint is so specific to GCs that it has been proposed as the
benchmark to classify a stellar system as a globular cluster
\citep{Car10c}.  Yet, neither the population as a whole, nor the two
sub-components of Terzan 5 show the Al-O anticorrelation.  Moreover,
each of the two populations shows spreads ($\sim 0.1$ dex) in both
[O/Fe] and [Al/Fe] not exceeding the 1$\sigma$ measurement errors (see
Table~\ref{tab2}), again at odds with the relatively large (several
tenths of dex) cosmic spreads measured in GCs of any metallicity
\citep{gra04}.

Hence, a second important result from our study is that Terzan~5
experienced a chemical enrichment history which is different from the
path typically leading to the development of the Al--O
anti-correlation and the presence of multiple stellar populations in
GCs \citep[see also][]{dercole08,dercole10} \footnote{However, note
  that while the most metal-rich GC where anti-correlations have been
  searched and found is NGC 6388 at [Fe/H]$=-0.4$ \citep{Car09}, both
  the populations of Terzan~5 lie in a metallicity range which has no
  direct counterpart among Galactic GCs.}.  In this respect, it is
also interesting to note that the stars in $\omega$ Centauri, while
sharing with Terzan~5 a significant spread in iron, clearly exhibit
anti-correlation signatures, both as a whole, and within each
metallicity sub-group, the only possible exception being the most
metal-rich stars \citep{JoPila}.  This may suggest that the chemical
enrichment history of Terzan~5 differs also from the one of $\omega$
Centauri.

Finally, Tables~\ref{tab1} and \ref{tab2} show that [C/Fe] is depleted
with respect to the Solar value in both populations. Such a carbon
depletion is commonly measured in the bulge giants \citep[see
  e.g.][]{rich07,ori08} and it indicates that some extra-mixing
processes are at work during the evolution along the RGB even at
metallicities close to Solar.

\section{Conclusions}

The main observational evidences from the photometric and
spectroscopic studies performed so far on Terzan 5 can be summarized
as follows.

\begin{itemize} 
\item Terzan~5 shows at least two stellar populations (as traced by
  both Red Clump and RGB stars) with distinct iron content and
  $[\alpha$/Fe] abundance patterns.  The metal-poor population
  ([Fe/H]$\simeq$-0.2) is $\alpha$-enhanced and closely resembles the
  bulk of the old bulge population (except for the extremely small
  spread in iron), which formed early and quickly from a gas mainly
  polluted by type II SNe.  The metal-rich population   has a
    metallicity ([Fe/H]$\simeq$+0.3), and an approximately
    scaled Solar [$\alpha$/Fe] ratio, requiring a progenitor gas
  further polluted by both type II and type Ia SNe on a longer
  timescale. It is difficult to place the chemistry of Terzan 5 within
  the framework of known GCs.  Indeed, while no genuine
  Galactic GC  displays
    such a large difference in the iron content, and even remotely
    resembles the metallicity regime of the two stellar populations
    of Terzan~5, stars
    with similar iron content have been observed in the bulge field
 \citep[see][]{rich07,ful07,zoc08}.

\item Neither Terzan~5 as a whole, nor the two populations separately
  show evidence of the Al-O anti-correlation.  As soon as the
  anti-correlation is also effective at Solar metallicity and above,
  this further suggests that Terzan~5 as a whole is not a genuine GC,
  and also that it cannot be the merging of two globulars.

\item Its current mass of a few $10^6 M_{\odot}$ \citep{lan10} is not
  sufficient to retain the SN ejecta and the large population of
  neutron stars which, thanks to an exceptionally high stellar
  collision rate \citep{verhut87,lan10}, could have been recycled into
  the multitude of MSPs that we observe today.
\end{itemize}

In order to draw more firm conclusions about the origin of Terzan~5
and its possible bimodal nature it is necessary to {\it (i)} complete
the chemical screening of its populations, by also sampling stars
that, in the CMD, are located between the two main RGBs, {\it (ii)}
perform and analyze ultra-deep IR imaging to accurately measure the
luminosity of the main sequence turn-off point(s) and derive the ages
of each component, and {\it (iii)} combine radial velocity and proper
motion measurements to properly determine the kinematics of the
system.  However, considering the information available so far, we
venture the following speculations.

The complex stellar population of Terzan~5 and the higher central
concentration of the most metal-rich component \citep[see F09
and][]{lan10} could be naturally explained within a self-enrichment
scenario.  An originally more massive proto-Terzan experienced the
explosions of a large number of type II and type Ia SNe, whose ejecta
have been retained within the potential well and which could also have
wiped out the anti-correlation signatures typical of GCs.  In such a
scenario, the Terzan~5 evolution should have been characterized by two
main and relatively short episodes of star formation, thus accounting
for the small metallicity spread of both populations.

In addition, the striking chemical similarity between Terzan~5
and the bulge population can also suggest a strong evolutionary
link between these two stellar systems and possibly a common origin
and evolution.  The current view \citep{kor04,imm04,she10} for the
formation of a bulge structure suggests a range of physical processes
that can be grouped in two main scenarios: (1) rapid formation
occurring at early epochs (as a fast dissipative collapse, mergers of
proto-clouds/sub-structures, evaporation of a proto-disk, etc.),
generating a spheroidal bulge populated by old stars, and (2)
evolution of a central disk/bar and its possible interaction with
other sub-structures on a longer timescale.  Within this framework,
Terzan 5 might well be the relic of a larger sub-structure that lost
most of its stars, probably because of strong dynamical interactions
with other similar systems at the early epoch of the Galaxy formation,
and/or later on with the central disk/bar. 
While most of the early fragments dissolved/merged together to form
the bulge, for some (still unclear) reasons Terzan 5 survived the
total disruption.  
Note that within this scenario, while the oldest
population of Terzan 5 would trace the early stages of the bulge
formation, the younger one could contain crucial information on its
more recent chemical and dynamical evolution.  The metal rich
sub-component of Terzan 5 stands as a remarkable stellar population,
worthy of more study.

\acknowledgements This research was supported by the Istituto
Nazionale di Astrofisica (INAF under contract PRIN-INAF 2008) and by
the Ministero dell'Istruzione, Universit\`a e Ricerca (MIUR). The
contribution of the Agenzia Spaziale Italiana (ASI, under contract
ASI/INAF I/009/10/0) is also acknowledged.  R.M. Rich acknowledges
support from grant AST 07-09479 from the National Science Foundation.
The authors wish to recognize and acknowledge the very significant 
cultural role and reverence that the summit of Mauna Kea has always 
had within the indigenous Hawaiian community.  We are most fortunate 
to have the opportunity to conduct observations from this mountain.

\newpage

\begin{deluxetable}{lllclcrllllllc}
\tabletypesize{\scriptsize} \tablecaption{Stellar parameters and
  abundances for the sample of observed giants in Terzan~5.}
\tablewidth{0pt} \tablehead{ \colhead{\#} & \colhead{RA}&
  \colhead{Dec}& \colhead{$\rm T_{eff}$}& \colhead{log~g}&
  \colhead{$\rm v_r^a$}& \colhead{$\rm [Fe/H]$}& \colhead{$\rm
    [O/Fe]$}& \colhead{$\rm [Si/Fe]$}& \colhead{$\rm [Mg/Fe]$}&
  \colhead{$\rm [Ca/Fe]$}& \colhead{$\rm [Ti/Fe]$}& \colhead{$\rm
    [Al/Fe]$}& \colhead{$\rm [C/Fe]$} } \startdata
1   &267.0182526 &-24.7787234 & 3400 & 0.5   & -85   & 0.18  &-0.06  &-0.03  & 0.12  & 0.05  & 0.02  & 0.12  &-0.28  \\
    &&&&&  &$\pm$0.05  &$\pm$0.10  &$\pm$0.16  &$\pm$0.14  &$\pm$0.09  &$\pm$0.14  &$\pm$0.17  &$\pm$0.09\\
2   &267.0176110 &-24.7777213& 3600 & 0.5   &-101    & 0.32  &-0.04  &-0.02  & 0.08  &-0.05  & 0.08  & 0.27  &-0.52  \\
    &&&&&  &$\pm$0.05  &$\pm$0.08  &$\pm$0.16  &$\pm$0.13  &$\pm$0.08  &$\pm$0.13  &$\pm$0.17  &$\pm$0.09\\
\enddata                       
\tablenotetext{a}{Heliocentric radial velocity in $\rm km~s^{-1}$.}
\tablecomments{The full version of the table is available in electronic form.}
\label{tab1}
\end{deluxetable}

\begin{deluxetable}{lcc}
\tablecaption{Average abundance ratios of the two RGB populations in Terzan~5.}
\tablewidth{0pt}
\tablehead{
\colhead{Abundance ratio} &
\colhead{metal-poor population}&
\colhead{metal-rich population}
}
\startdata
$\rm [Fe/H]$  & $-0.25 \pm 0.07$ & $+0.27 \pm 0.04$ \\ 
$\rm [O/Fe]$  & $+0.34 \pm 0.06$ & $-0.04 \pm 0.04$ \\
$\rm [Ca/Fe]$ & $+0.32 \pm 0.05$ & $+0.02 \pm 0.03$ \\ 
$\rm [Si/Fe]$ & $+0.36 \pm 0.08$ & $+0.02 \pm 0.10$ \\
$\rm [Mg/Fe]$ & $+0.33 \pm 0.10$ & $+0.08 \pm 0.06$ \\
$\rm [Ti/Fe]$ & $+0.34 \pm 0.10$ & $+0.06 \pm 0.06$ \\
$\rm [Al/Fe]$ & $+0.52 \pm 0.13$ & $+0.13 \pm 0.13$ \\
$\rm [C/Fe]$  & $-0.35 \pm 0.12$ & $-0.38 \pm 0.08$ \\
\enddata                       
\label{tab2}
\end{deluxetable}

\end{document}